# Database Technology Evolution III: Knowledge Graphs and Linked Data

Malcolm Crowe
Emeritus Professor, Computing Science
University of the West of Scotland
Paisley, United Kingdom
Email: Malcolm.Crowe@uws.ac.uk

Fritz Laux
Emeritus Professor, Business Computing
Reutlingen University
Reutlingen, Germany
Email: Fritz.Laux@reutlingen-university.de

*Abstract*— This paper reviews the changes for database technology represented by the current development of the draft international standard ISO 39075 (Database Languages - GQL), which seeks a unified specification for property graphs and knowledge graphs. This paper examines these current developments as part of our review of the evolution of database technology, and their relation to the longer-term goal of supporting the Semantic Web using relational technology.

## I. INTRODUCTION

Tim Berners-Lee originated the concept of the Semantic Web in 1999, as a way of enabling computers to analyze all the content, links and transactions between people and computers on the Web [1]. Initial approaches to this dream focused initially on the addition of semantic information to everything in all documents [2], documenting semantic information using subject-relation-object triples. Thinking of objects as nodes or vertices and triples as edges or relationships yields the concept of a knowledge graph [3][4][5]. While some triples merely described the content in a document, those that were links to other documents proved to be more interesting to human readers, leading to the topic of linked data [6]. There are now many open data projects whose nodes are items of information on the Web with less focus today on the detail of document internals [7].

The underlying technology for managing such knowledge bases originally seemed completely different from relational databases, which processed representations in the form of tabular data while in knowledge bases the links were first-class objects. There were also differences in scale: databases dealt with the needs of individual companies, while knowledge is worldwide.

Graph database technology is more efficient than relational technology in following chains of relationships, because in relational technology such sequences imply joins of all the corresponding tables. Many graph database products are now available [8] and the business case for further development in this area is compelling, with use cases including medical research [9], fraud detection [10] and cybersecurity [11], global engineering design [12] and supply chain management [13].

Even the most radical products for processing knowledge data use data storage, and there is now a new international standard for a database language GQL [14] to include triple graphs and property graphs (the name GQL in the title of the standard is not an acronym, although some authors have been persuaded to invent a three word phrase with these initials). In the past, databases of triples (subject, predicate, object) tied to HTTP urls looked very different from databases consisting of linked sets of objects with given property values. Implementations of this new GQL standard can be expected soon.

In 2023 we reported at IARIA [15, sec III.A] on a way of implementing GQL by adding new metadata to the ISO 9075 Standard Query Language (SQL) [16], and we exemplified this in 2024 with a brief account of a Financial Benchmark for GQL [17][18]. The implementation described was relational in nature, using the property graph approach of GQL, and did not discuss knowledge graphs, cross-platform linked data, issues evident in the recent research papers referenced above, so that it makes sense to continue our story of database evolution [19][20] in this paper by introducing a very lightweight implementation of knowledge graphs and web services.

This paper is thus a practical contribution to data and systems research, through concept development in the context of a lightweight open-source proof of concept implementation [21]. It also takes up the question of semantic alignment from [22] and implements ideas for graph schema under discussion in the GQL community.

The plan of this paper is to motivate these developments in Section II, with the help of two examples from recent publications and some discussion of related implementation issues. The first example, in Figure 1, is from [23] and links two graphs, the second, in Table 1, is from [5] and illustrates the triples approach to knowledge representation. We briefly cover linking data by web services in Section III, and graph schema ideas in Section IV. Section V provides some conclusions and our plans for completing the work as an open-source research contribution.

## II. KNOWLEDGE GRAPH IMPLEMENTATION

Neither example in this section fits well with relational database model, and they continue to require development of the GQL specification.

### A. The Yacht Club example

The current edition of GQL allows "open" graphs without defined graph types and "closed" graphs where all node and edge types are predefined, but there is no mechanism for modifying such types once defined.

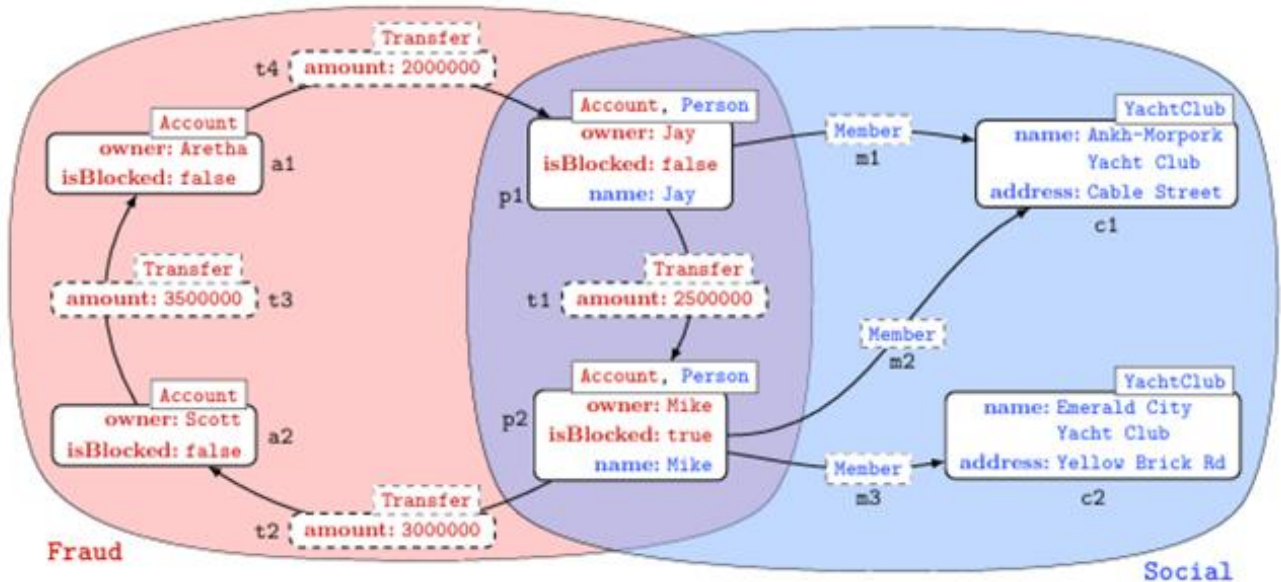

Figure 1: The Yacht Club example [23]

TABLE I: CREATING THE GRAPH OF FIGURE 1 IN GQL

```
create schema /yc;
create graph type /yc/Social {node Person {name string},
        node YachtClub {name string,address string},
        directed edge "Member" connecting (Person->YachtClub)};
create graph /yc/Fraud ANY;
insert (a2 :Account{owner:'Scott',isBlocked:false})-[:Transfer{amount:350000}]->
(:Account{owner:'Aretha',isBlocked:false})-[:Transfer{amount:2000000}]->
(p1 :Person&Account{owner:'Jay',name:'Jay',isBlocked:false})
-[:"Member"]->(:YachtClub {name:'Ankh-Morpork Yacht Club',address: 'Cable Street'})
<-[:"Member"]-(p2 :Person&Account {owner:'Mike',name:'Mike', isBlocked:true})
-[:"Member"]->(:YachtClub{name:'Emerald City Yacht Club',address:'Yellow Brick Road'}),
(p1)-[:Transfer{amount:2500000}]->(p2)-[:Transfer{amount:3000000}]->(a2);
```

However, this example is motivated by combining information from two separately developed graphs.

In Figure 1, we see two graphs called Fraud and Social, both of which contain nodes p1 and p2. In the Fraud graph, these are of type Account, while in the Social graph, they are of type Person. Nodes cannot belong to several graphs in the current GQL standard, and GQL statements can make changes to the data in at most one graph. With a little goodwill on these points, the script in Table I constructs an open graph (Fraud) and a closed graph type (Social).

Here the labels Person and Account make up p1's *label set*. GQL types have label sets (unlike its predecessors such as Neo4j). The single node p1 has properties from a node type in each graph and so belongs to both graphs, while Person&Account is a label expression, not a node type. From the relational database viewpoint, tables consist of relations of the same type, so that, if both Person and Account are row types, each corresponding table gets a row when the record for p1 is inserted.

Open graphs allow new node and edge types to be introduced on insertion, but labels such as Person and Account need to be well defined (property sets, connections) before they can be combined with others. In Table II, Transfer is defined as connecting Account nodes before it is used for Person&Account. Note that the aliases a2, p1, and p2 are local to the insert statement as is usual in SQL. Using match-insert combinations as suggested above can avoid long insert sequences.

The second example is shown in Table II.

It needs to declare somehow that $<_{sp}$ is a relation between edge types. This also would require changes to the current edition of the GQL standard and we return to this point in Section IV below.

TABLE II: A KNOWLEDGE GRAPH [5]

$t_1$ = (:John :masterFrom :DauphineUni),
$t_2$ = (:John :phdFrom :DauphineUni),
$t_3$ = (:masterFrom $<_{sp}$ :degreeFrom),
$t_4$ = (:phdFrom $<_{sp}$ :degreeFrom)

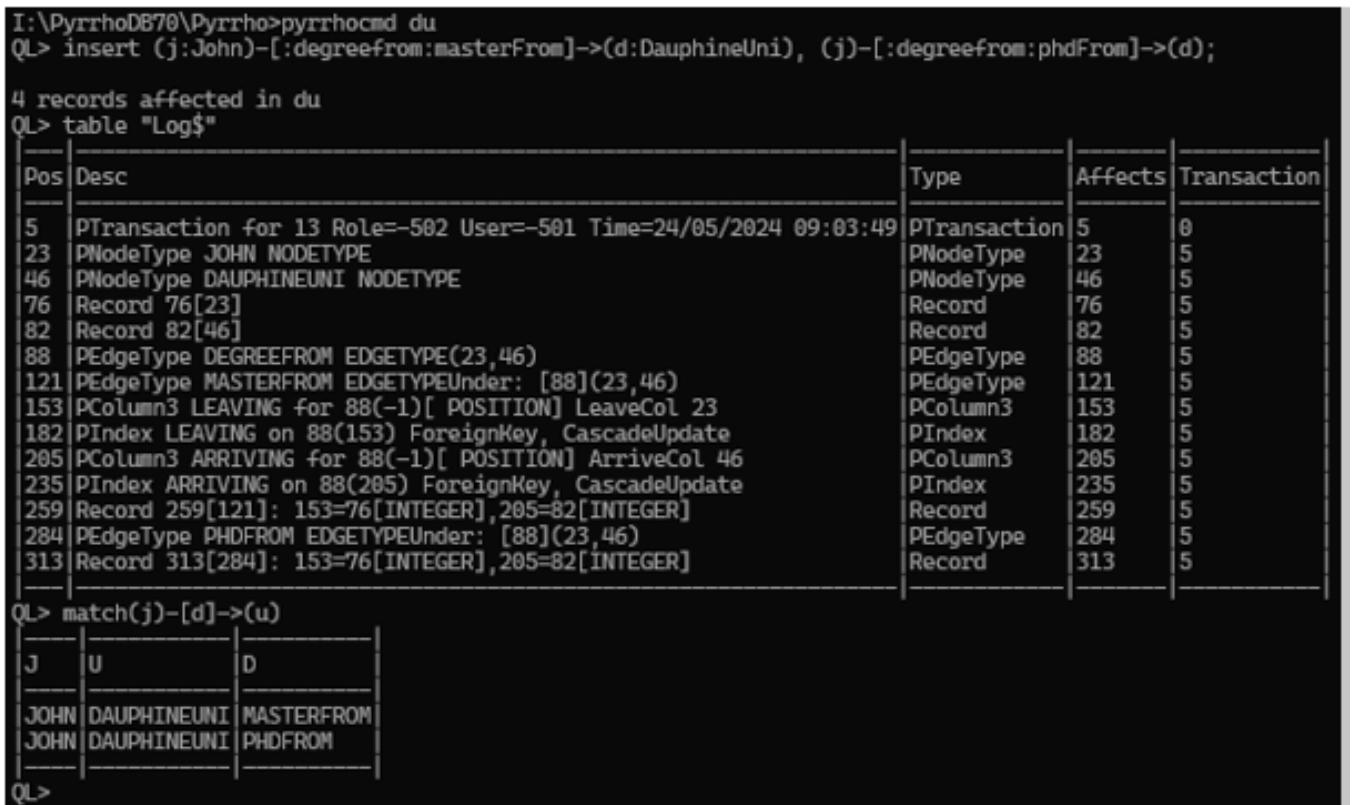

Figure 2: The transaction log and a simple Match statement for the example in Table II.

In [15, III.C] we showed that graphs based on SQL user defined types can be constructed without prior declaration of types (in GQL this is called using open graph types), so the simple database for the second example can be constructed in Pyrrho with just one statement. Starting with an empty database,

insert (j:John)-[:degreefrom:masterFrom]->
(d:DauphineUni), (j)-[:degreefrom:phdFrom]->(d);

Figure 2 shows the transaction log resulting from this statement in Pyrrho: it shows the mixture of type and data creation steps This little database occupies only 339 bytes on disk.

Match statements in GQL provide a simple way of retrieving information from a graph, by binding free variables to graph contents according to a graph pattern. A graph pattern can specify labels or properties required for the match: in this case there is no need to do so. Graph patterns can also specify alternatives and trails through the graph, and Match can have dependent statements with access to the binding results, such as selection (or RETURN) of results and aggregations, or data-modifying statements such as INSERT, DELETE, SET or REMOVE that can modify the graph and its contents.

### B. An open-source prototype GQL implementation

Our sample implementation, Pyrrho [21] has the ambition not only to address both these examples together with GQL and SQL syntax. By design, the GQL specification has chosen to accommodate this sort of fusion, but there is an issue that some of SQL's reserved words are not reserved in GQL. This will mean that if a database defines some SQL reserved words to mean something else, syntax depending on these words will not be available.

As a research laboratory for database management, Pyrrho also has been evolving for over two decades, and as of May 2024 it accommodates graph objects (node and edges) and their types alongside the standard SQL apparatus, in the manner described above.

Some basic features of Pyrrho make the task of implementing GQL easier. First, in this RDBMS the database file is an append-storage transaction log so that the position of any committed database object or record does not change even if the contents are updated. Pyrrho makes this position into a pseudo-column, so that the next step in the evolution of our implementation is to use this position where primary keys would normally be used. Introducing this sort of flexibility into a relational DBMS is quite a step.

Another feature of Pyrrho is its optimistic concurrency control based on shareable data structures. These two aspects allow transactions to mix schema changes and data modification and avoid the complications involved in two-phase locking.

Pyrrho already provides triggers and type alteration. GQL's structure comes from the edge relationships, so that the current edition of GQL does not have any concept of integrity constraints such as primary keys or foreign keys. Many business applications can benefit from the additional structure provided by allowing relational constraints in a graphical database.

In this section it remains to include a brief discussion of the effect on the database model of Pyrrho [21]. Before the evolution above, the node and edge structure of graph database models used primary and foreign keys. so that columns ID, LEAVING and ARRIVING would be added to node types and edge types, and values for these would be added if they were not provided. From the viewpoint of GQL, this process is unnecessary, and now in Pyrrho the position pseudocolumn is used instead of a new ID column, but if an Id or primary key is already in a new node type it will be used instead. The metadata syntax for declaring node types and edge types includes ways of specifying which existing columns are used for such structural properties.

This leads to smaller and faster implementation of large graphs: smaller because fewer indexes need to be constructed or checked, and faster because the overhead of finding suitable values for the automatic keys is not required.

Pyrrho's client program currently requires multiline statements to be enclosed in square brackets, so that square brackets within multiline statements should not be at the ends of lines.

### III. WEB SERVICES VS BIG DATA

This section describes an implementation of cross-platform linking of data. Previous work [24] discussed how data distributed in different institutions could be processed without the mass import of linked data by extra-transform-load. The key idea was view mediation: a view could be defined with a url for retrieval from a remote source. Assuming that the remote source granted the necessary authorizations, selection and modification of remote data could be allowed using HTTP, and with HTTP POST the mechanism allowed a sequence of operations to be performed on the remote system in a single transaction.

GQL has no details yet on viewed graphs, but the basic idea is clear: like a viewed table, the system can retrieve and process, but does not store, the viewed contents.

A suitable syntax for supplying the url for such remote access is in GQL's USE GRAPH syntax. We can simply write USE GRAPH (url) . As before it is up to the remote system to grant access: the local system will provide its CURRENT_USER information within the HTTP header. Ordinary GQL statements follow the USE GRAPH and become the body of the HTTP POST request, and the result of the final step (e.g. MATCH or RETURN) will be returned from the remote server, along with a suitable ETag as described in [24].

### IV. GRAPH SCHEMA IMPLEMENTATION

A new suggestion for Graph Schema has arisen in discussions about GQL [25] that is very close to the suggestions for Typed Graph Schema in [21]. The idea is that for any graph G, the Graph Schema should itself have the form of a graph S so that nodes of S are node types of G, edges of S are edge types of G, the properties of object types in S are the property types of corresponding objects in G.

Schema information can then be accessed using a MATCH SCHEMA statement. The vision here is that data-modifying statements affecting S should provide a mechanism for altering the graph types of G.

For example, it could be argued that since [5] is all about the consequences of implication, the discussion of example 2 above assumed assertions t3 and t4 (see Table II) at the outset. It would be more in keeping with the context of [5] to be able to implement example 2 using the original triples as follows:

```
INSERT (:John)-[:masterFrom]->(:DauphineUni);
INSERT (:John)-[:phdFrom]->(:DauphineUni);
INSERT SCHEMA [:masterFrom=>:DegreeFrom];
INSERT SCHEMA [:phdFrom=>:DegreeFrom];
```

This respects the statement in [5] that the last two triples specify a relationship at the schema level: that :masterFrom and :phdFrom are subproperties of :degreeFrom, and we assume this is done without creating new instance nodes in the graph. In an open graph, the first statement would be allowed in GQL, with the use of unbound identifiers in the first INSERT implying the creation of nodes and edges and associated (singleton) types. The second does something similar for the unbound label :phdFrom but is at least unusual in its use of the singleton labels :John and :DauphineUni. Normally such an insertion would be specified by a statement such as

```
MATCH (j:John),(u:DauphineUni)
INSERT (j)-[:phDFrom]->(u);
```

If so, it is arguable that the third statement implies the creation of an edge type for the unbound :DegreeFrom, while the fourth inserts the implies relationship.

This represents ongoing research in discussions with the GQL community.

### V. CONCLUSIONS

This short paper has provided some notes on the current developments in the new database language GQL, and their relationship with recent research papers on knowledge graphs and linked data. Our SQL implementation, Pyrrho [21] is being updated to take account of these changes, and in time will implement all of GQL. Despite this ambition, Pyrrho's executable binaries are very lightweight (less than 2 MB in total) and are very economical with disk space as indicated in section II above. Pyrrho's test suite includes simple cases that show the integration of the relational and typed graph model concepts, and benchmark tests on databases of 500MB show that the design scales well.

Research will continue in order to find the best way of implementing a full GQL implementation while offering a full SQL feature set.